\documentclass[11pt]{article}

\usepackage{amssymb,amsmath,amsthm}
\usepackage{graphicx}
\usepackage{latexsym}

\newcommand{\NP}{\ensuremath{ {\rm NP}}}
\renewcommand{\P}{\ensuremath{ {\rm P}}}
\newcommand{{\Nat}}{{\rm I\! N}}
\newcommand{{\BF}}{{\rm BF}}
\newcommand{{\Forb}}{{\rm Forb}}
\newcommand{{\pot}}{{\mathcal P}}
\newcommand{{\F}}{{\mathcal F}}
\newcommand{{\D}}{{\mathcal D}}
\newcommand{{\BFD}}{{\rm D}}
\newcommand{{\BFR}}{{\rm R}}
\newcommand{{\BFM}}{{\rm M}}
\newcommand{{\BFL}}{{\rm L}}
\newcommand{{\XC}}{{\mathcal C}}
\newcommand{{\XS}}{{\mathcal S}}
\newcommand{{\XD}}{{\mathcal D}}
\newcommand{{\XG}}{{\mathcal G}}
\newcommand{{\XF}}{{\bf F}}
\newcommand{{\XX}}{{\mathcal X}}

\newcommand{{\sd}}{{\mathrm{sd}}}
\newcommand{{\id}}{{\mathrm{id}}}
\newcommand{{\nt}}{{\mathrm{not}}}
\newcommand{{\dual}}{{\mathrm{dual}}}
\newcommand{{\CSP}}{{\mathrm{CSP}}}
\newcommand{{\lookup}}{{\mathrm{T}}}
\newcommand{{\formula}}{{\mathrm{F}}}
\newcommand{{\circuit}}{{\mathrm{C}}}
\newcommand{{\AND}}{{\mathrm{AND}}}
\newcommand{{\OR}}{{\mathrm{OR}}}
\newcommand{{\XOR}}{{\mathrm{XOR}}}
\newcommand{{\XNOR}}{{\mathrm{XNOR}}}

\newcommand{{\FPE}}{{\textsc{FixedPoints}}}
\newcommand{{\SAT}}{{\textsc{3SAT}}}
\newcommand{{\PSAT}}{{\textsc{Planar 3SAT}}}
\newcommand{{\PHSAT}}{{\textsc{Planar Horn-2SAT}}}
\newcommand{{\POSSAT}}{{\textsc{Pos 2SAT}}}

 \newtheorem{theorem}{Theorem}[section]
 \newtheorem{corollary}[theorem]{Corollary}
 
 \newtheorem{proposition}[theorem]{Proposition}
 \theoremstyle{definition}
 
 \theoremstyle{remark}

 \numberwithin{equation}{section}

\begin{document}

\title{Dichotomy Results for Fixed-Point Existence Problems for Boolean Dynamical Systems}

\author{{\em Sven Kosub}\\
Fakult\"at f\"ur Informatik, Technische Universit\"at M\"unchen, \\
Boltzmannstra{\ss}e 3, D-85748 Garching, Germany\\
{\tt kosub@in.tum.de}}

\date{\empty}

\maketitle

\begin{abstract}
A complete classification of the computational complexity of the 
fixed-point existence problem for boolean dynamical systems, i.e., finite 
discrete dynamical systems over the domain $\{0,1\}$, is presented. 
For function classes $\F$ and graph classes $\XG$, an $(\F,\XG)$-system is a 
boolean dynamical system such that all local transition functions lie in $\F$ 
and the underlying graph lies in $\XG$. 
Let $\F$ be a class of boolean functions which is closed under composition 
and let $\XG$ be a class of graphs which is closed under taking minors. 
The following dichotomy theorems are shown:
(1) If $\F$ contains the self-dual functions and $\XG$ contains the planar 
	graphs, then the fixed-point existence problem for $(\F,\XG)$-systems with 
	local transition function given by truth-tables is $\NP$-complete; otherwise, 
	it is decidable in polynomial time.
(2) If $\F$ contains the self-dual functions and $\XG$ contains the graphs 
	having vertex covers of size one, then the fixed-point existence problem for 
	$(\F,\XG)$-systems with local transition function given by formulas or 
	circuits is $\NP$-complete; otherwise, it is decidable in polynomial time.
\end{abstract}

\maketitle

\section{Introduction}

{\em Background on complex systems.} 
A complex system, in a mathematical sense, can be viewed as a collection
of highly interdependent variables. A discrete dynamical system is a complex 
system where variables update their values in discrete time. 
Though the interdependencies among the variables might have quite simple 
descriptions on a local level, the overall global behavior of the systems 
can be as complicated as unpredictable or undecidable 
(see, e.g., \cite{buss-papadimitriou-tsitsiklis-1991,moore-1990}).
This phenomenon has been widely studied in the theory of cellular automata 
\cite{neumann-1966,wolfram-1994} and its applications (see, e.g., \cite{garzon-1995,chattopadhyay-chaudhuri-chowdhury-nandi-1997}).

Finite discrete dynamical systems are characterized by finite sets of variables
which can take values from a finite domain.
In essence (see, e.g., \cite{barrett-reidys-1999,barrett-mortveit-reidys-2000}), a 
finite discrete dynamical system (over a finite domain) consists of 
(1) a finite undirected graph, where vertices correspond to variables and edges 
correspond to an interdependence between the two connected variables, 
(2) a set of local transition functions, one for each vertex, that map values 
of variables depending on the values of all connected variables 
to new variable values, and 
(3) an update schedule that governs which variables are allowed to update their 
values in a certain time step. 
A formal definition is given in Sect.~\ref{sec:dynamical-systems}.
Due to their structural simplicity and modelling flexibility, finite discrete 
dynamical systems are suitable for analyzing the behavior of real-world 
complex systems. In fact, the conception is motivated by analysis and simulation 
issues in traffic flow (see, e.g, \cite{barrett-bush-kopp-mortveit-reidys-2000,
barrett-eubank-marathe-mortveit-reidys-2000}) and inter-domain routing 
\cite{griffin-wilfong-1999}.
It also has applications to two-species diffusion-reaction systems such as 
synchronous or asynchronous versions of the nearest neighbor coalescence reaction
$A+A\to A$ on a lattice in the immobile reactant case \cite{abad-grosfils-nicolis-2001}.

A central problem in the study of discrete dynamical systems is the 
classification of systems according to the predictability of their behavior. 
In the finite setting, a certain behavioral pattern is considered predictable if 
it can be decided in polynomial time whether a given system will show the 
behavioral pattern \cite{buss-papadimitriou-tsitsiklis-1991}. 
In a rather strong sense, predictability and tractability are identified. 
It is not surprising that the reachability of patterns is, in general, an 
intractable problem, i.e., at least 
$\NP$-hard (see, e.g., \cite{green-1987,sutner-1995,
barrett-hunt-marathe-ravi-rosenkrantz-stearns-2006}). 
However, some restricted subclasses of finite discrete dynamical systems, i.e., 
systems given by restricted sets of local transition functions and 
network topologies, are known to possess easy-to-predict patterns (see, e.g., 
\cite{barrett-hunt-marathe-ravi-rosenkrantz-stearns-2003,
barrett-hunt-marathe-ravi-rosenkrantz-stearns-2003a,
barrett-hunt-marathe-ravi-rosenkrantz-stearns-2006} and the discussion of 
related work below). 
For the purpose of analyzing and simulating real-world systems by finite discrete 
dynamical systems, it is highly desirable to have sharp boundaries
between tractable and intractable cases.

A fundamental behavioral pattern for discrete dynamical systems are fixed points 
(homogeneous states, equilibria). 
A value assignment to the variables of a system is a fixed point if the values 
assigned to the variables are left unchanged if the system updates the values. 
A series of recent papers has been devoted to identification of finite systems with 
tractable/intractable fixed-point analyses 
\cite{barrett-hunt-marathe-ravi-rosenkrantz-stearns-tosic-2001,agha-tosic-2005,
tosic-2005,tosic-2006}.
But although it is an old question how common intractability results are for 
discrete dynamical systems (see \cite[Problem 19]{wolfram-1985}), a 
precise characterization of the {\em islands of tractability} even for the 
fixed-point existence problem in the simplest case of the boolean domain $\{0,1\}$ 
has remained an open problem 
(see \cite{barrett-hunt-marathe-ravi-rosenkrantz-stearns-tosic-2001}).

\medskip

\noindent
{\em Contributions of the paper.}
We contribute to the problem of classifying boolean (discrete) dynamical systems,
i.e., finite discrete dynamical systems over the domain $\{0,1\}$, with regard to 
the computational complexity of the fixed-point existence problem, i.e., decide
whether a given system has a fixed point, in two ways.

A first contribution is the proposal of a general analysis framework for systems. 
We say that a boolean dynamical system is an $(\F,\XG)$-system if its local 
transition functions belong to the class $\F$ of boolean functions and the 
underlying graph belongs to the graph class $\XG$. 
We propose to consider two well-studied frameworks for functions and graphs (a 
formal introduction is given in Sect.~\ref{sec:analysis-framework}):
\begin{itemize}
\item  {\em Post classes} \cite{post-1941}: We assume that the function class $\F$ 
	is closed under composition (and some more reasonable operations). Equivalently,
	$\F$ is a class of boolean functions that can be built by arbitrary circuits 
	over gates from some finite logical basis \cite{post-1941}. Examples are the 
	monotone functions, the linear functions, and the self-dual functions (i.e., 
	functions $f$ such that $f(x_1,\dots,x_n)=1-f(1-x_1,\dots,1-x_n)$).
\item {\em Graph minor classes} (see, e.g., \cite{diestel-2003}): We assume that
	the graph class $\XG$ is closed under taking minors, i.e., $\XG$ is closed 
	under vertex deletions, edge deletions, and edge contractions. Equivalently, 
	$\XG$ can be characterized by a finite set of forbidden minors 
	\cite{robertson-seymour-2004}. An example is the class of planar graphs 
	(together with the forbidden minors $K_{3,3}$ and $K^5$).
\end{itemize}
\noindent
Certainly, other schemes can be devised for system classifications. In fact, many 
results in literature do not fit into this framework (see the discussion of related 
work below). However, the proposed scheme has strong features: first, it exhausts the 
class of all boolean dynamical systems; second, for fixed function classes $\F$ and 
fixed graph classes $\XG$, it is decidable in polynomial time whether a given system 
is in fact, an $(\F,\XG)$-system (supposed the local transition functions are 
represented by lookup-tables); and third, it allows elegant proofs of dichotomy 
theorems (as exemplified by our second contribution).
We mention that, as fixed points are invariant under changes of the update regime, 
a scheme for classifying update schedules is not needed for our study.

The main contribution of the paper is a complete complexity classification of
the fixed-point existence problem with respect to our analysis framework. 
We make a distinction of the problem into three categories: systems for which 
the local transition functions are given by lookup-tables, by formulas (over 
logical bases), or by circuits (over logical bases). 
For each case we obtain a tractability/intractability dichotomy.
Interestingly, the dichotomy theorems for formulas and circuits coincide.
Let $\F$ be a Post class of boolean functions and let $\XG$ be graph class 
closed under taking minors. The following is proved in Sect.~\ref{sec:islands}:
\begin{itemize}
\item {\em Dichotomy for boolean dynamical systems based on lookup-tables}:
	If $\F$ contains 
	the self-dual functions and $\XG$ contains the planar graphs, then the 
	fixed-point existence problem for $(\F,\XG)$-systems with 
	local transition functions given by lookup-tables is $\NP$-complete; in 
	all other cases, it is decidable in polynomial time 
	(Theorem \ref{thm:fpe-dichotomy-t}).
\item {\em Dichotomy for boolean dynamical systems based on formulas or circuits}: 
	If $\F$ contains the self-dual functions and 
	$\XG$ contains the graphs having vertex covers of size one, then the 
	fixed-point existence problem for $(\F,\XG)$-systems with local transition 
	functions given by formulas/circuits (over the logical basis of $\F$) is 
	$\NP$-complete; in all other cases, it is decidable in polynomial time
	(Theorem \ref{thm:fpe-dichotomy-f}).
\end{itemize}
The results provide easy criteria for deciding whether $(\F,\XG)$-systems have 
tractable or intractable fixed-point existence problems. For instance, tractability
follows for systems with linear or monotone local transition function on arbitrary networks 
(see also \cite{barrett-hunt-marathe-ravi-rosenkrantz-stearns-tosic-2001}). The tractability 
regions with respect to arbitrary local transition functions and restricted graphs 
correspond to bounded treewidth (in the case of lookup-tables) and bounded degree 
(in the case of formulas or circuits). Tractable network classes (for 
lookup-table-based systems) are, e.g., trees, outer-planar graphs, or series-parallel 
graphs. 

\medskip

\noindent
{\em Related work.} 
There is a series of work regarding the complexity of certain computational 
problems for discrete dynamical systems (see, e.g., 
\cite{green-1987,sutner-1995,barrett-hunt-marathe-ravi-rosenkrantz-stearns-tosic-2001,
barrett-hunt-marathe-ravi-rosenkrantz-stearns-2003,
barrett-hunt-marathe-ravi-rosenkrantz-stearns-2003a,agha-tosic-2005,tosic-2005,
barrett-hunt-marathe-ravi-rosenkrantz-stearns-2006} and the
references therein). 

Detailed studies of computational problems related to fixed-point existence 
have been reported in \cite{barrett-hunt-marathe-ravi-rosenkrantz-stearns-tosic-2001,
agha-tosic-2005,tosic-2005,tosic-2006}. 
As shown in \cite{barrett-hunt-marathe-ravi-rosenkrantz-stearns-tosic-2001},
tractable cases for fixed-point existence are constituted by systems with 
linear, generalized local transition functions,
systems with monotone local transition 
functions and systems where local transition function are computed by gates 
(of unbounded fan-in) from $\{\text{AND, OR, NAND, NOR}\}$; intractable 
cases are boolean dynamical systems having local transition functions
computed by gates (of unbounded fan-in) from the sets $\{\text{NAND, XNOR}\}$, 
$\{\text{NAND, XOR}\}$, $\{\text{NOR, XNOR}\}$, or $\{\text{NOR, XOR}\}$. 
Moreover, in \cite{agha-tosic-2005,tosic-2005,tosic-2006}, the problem of 
enumerating fixed points of boolean dynamical systems has been studied: 
counting the number of fixed points is in general $\#\P$-complete, 
even counting the number of fixed points for boolean dynamical systems
with monotone local transition functions over planar bipartite graphs or
over uniformly sparse graphs is $\#\P$-complete. We note that all system classes
considered are based on formula or circuit representations, i.e., the 
intractability results fall into the scope of Theorem \ref{thm:fpe-dichotomy-f}.

Mainly, tractability and intractability results have been
shown for various version of pattern reachability problems such as 
garden of Eden existence (e.g., \cite{barrett-hunt-marathe-ravi-rosenkrantz-stearns-tosic-2001,agha-tosic-2005,tosic-2005}),
predecessor existence (e.g., \cite{sutner-1995,barrett-hunt-marathe-ravi-rosenkrantz-stearns-2003a}),
parameterized and unparameterized reachability 
(e.g., \cite{green-1987,sutner-1995,barrett-hunt-marathe-ravi-rosenkrantz-stearns-2003a, 
barrett-hunt-marathe-ravi-rosenkrantz-stearns-2003b,
barrett-hunt-marathe-ravi-rosenkrantz-stearns-2006}).
To summarize, the system subclasses considered restrict the local 
transition functions to linear functions, monotone functions, 
various types of threshold functions, or symmetric functions.
Except the linear and monotone functions none of these classes is closed under 
composition. Restrictions to the dependency networks involve planar graphs, 
regular graphs, bounded-degree graphs, star networks, and bounded pathwidth.

In the theory of finite discrete dynamical systems, tight dichotomy results 
have been shown for very restricted classes of systems (as, 
e.g., in \cite{buss-papadimitriou-tsitsiklis-1991}).
Exhausting results similar to those of this paper are standard for constraint
satisfaction problems (see, e.g., \cite{boehler-creignou-reith-vollmer-2003} 
for a survey). In fact, our results rely on close relationships between 
fixed-point and constraint satisfaction problems (as in the proof of Theorem 
\ref{thm:tractable-networks}).

\section{The Dynamical Systems Framework}
\label{sec:dynamical-systems}

In this section, we describe our model of dynamical systems. 
We follow the approach given by \cite{barrett-mortveit-reidys-2000} with a 
marginal generalization regarding update schedules. 

\medskip

\noindent
{\em Dynamical systems.} The underlying network structure of a dynamical 
system is given by an undirected graph $G=(V,E)$ without multi-edges and 
loops. We suppose that the set $V$ of vertices is ordered. So, 
without loss of generality, we assume $V=\{1, 2, \dots,n\}$. For any 
vertex set $U\subseteq V$, let $N_G(U)$ denote the neighbors of $U$ in $G$, 
i.e., 
\[N_G(U) =_{\rm def} \{~j~|~j\notin U\textrm{ and there is an $i\in U$ 
such that $\{i,j\}\in E$}~\}.\]
If $U=\{i\}$ for some vertex $i$, then we use $N_G(i)$ as a shorthand for 
$N_G(\{i\})$. Define $N^0_G(i)=_{\rm def} \{i\}\cup N_G(i)$. The degree 
$d_i$ of a vertex $i$ is the number of its 
neighbors, i.e., $d_i=_{\rm def} \|N_G(i)\|.$ 

A {\em dynamical system $S$ over a domain $\XD$} is a pair $(G, F)$ where 
$G=(V,E)$ is an undirected graph (the {\em network}) and $F=\{f_i~|~i\in V\}$ 
is a set of {\em local transition functions} $f_i:\XD^{d_i+1}\to \XD$.
The intuition of the definition is that each vertex $i$ corresponds 
to an active element (entity, agent, actor etc.) which is always in some 
state $x_i$ and which is capable to change its state, if necessary. The 
domain of $S$ formalizes the set of possible states of all vertices of the 
network, i.e., for all $i\in V$, it always holds that $x_i\in\XD$. A vector 
$\vec{x}=(x_i)_{i\in V}$ such that $x_i\in \XD$ for all $i\in V$ is called 
a {\em configuration of $S$}. If it is more convenient, then we also say 
that a mapping $I:V\to\XD$ is a configuration. A {\em subconfiguration} of 
$I$ with respect to $A\subseteq V$ is a mapping $I[A]:A\to\XD$ such that 
$I[A](i)=I(i)$ for all $i\in A$.
The local transition function 
$f_i$ for some vertex $i$ describes how $i$ changes its state depending on 
the states of its neighbors $N_G(i)$ in the network and its own state. 

\medskip
\noindent
{\em Discrete dynamical systems.} We are particularly interested in dynamical 
system operating on a discrete time-scale. A {\em discrete dynamical system 
$\XS=(S,\alpha)$} consists of a dynamical system $S$ and a mapping $\alpha:
\{1,\dots,T\}\to \pot(V)$, where $V$ is a set of vertices of the network of 
$S$ and $T\in\Nat$. The mapping $\alpha$ is called the {\em 
update schedule} and specifies which state updates are realized at certain 
time-steps: for $t\in\{1,\dots,T\}$, $\alpha(t)$ specifies those vertices that 
simultaneously update their states in step $t$. 

\medskip
\noindent
{\em Global maps.} A discrete dynamical system $\XS=(S,\alpha)$ over domain 
$\XD$ induces a global map $\XF_\XS:\XD^n \to\XD^n$ where $n$ is the number 
of vertices of $S$. For each vertex $i\in V$, define an {\em activity function} 
$\varphi_i$ for a set $U\subseteq V$ and $\vec{x}=(x_1,\dots,x_n)\in\XD^n$ by
\[
\varphi_i[U](\vec{x})=_{\rm def}\left\{\begin{array}{ll}
f_i(x_{i_1},\dots,x_{i_{d_i+1}})  &\textrm{ if $i\in U$}\\
x_i & \textrm{ if } i\notin U
\end{array}\right.\]
where $\{i_1,i_2,\dots,i_{d_i+1}\}=N^0_G(i)$. For a set $U\subseteq V$, 
define the {\em global transition function} $\XF_S[U]:\XD^n\to \XD^n$ for all 
$\vec{x}\in\XD^n$ by \[\XF_S[U](\vec{x})=_{\rm def} (\varphi_1[U](\vec{x}),
\dots,\varphi_n[U](\vec{x})).\] Note that the global transition function does 
not refer to the update schedule, i.e., it only depends on the dynamical system 
$S$ and not on $\XS$. The function $F_\XS:\XD^n\to\XD^n$ computed by the discrete 
dynamical system $\XS$, the {\em global map} of $\XS$, is defined by
\[\XF_\XS=_{\rm def} \prod_{k=1}^T \XF_S[\alpha(k)].\]

\noindent
{\em Fixed points.} The central notion for our study of dynamical systems is the 
concept of a fixed point, i.e., a configuration which does not change under any 
global behavior of the system.
Let $S=(G,\{f_i~|~i\in V\})$ be a dynamical system over domain $\XD$. 
A configuration $\vec{x}\in \XD^n$ is said to be a {\em local fixed 
point of $S$ for $U\subseteq V$} if and only if $\XF_S[U](\vec{x})=
\vec{x}$.
A configuration $\vec{x}\in \XD^n$ is said to be a {\em fixed point 
of $S$} if and only if $\vec{x}$ is a local fixed point of $S$ for $V$.

The following useful proposition is easily seen.
\begin{proposition} \label{prop:fundamental}
Let $S=(G,\{f_i~|~i\in V\})$ be a dynamical system over domain $\XD$. 
\begin{enumerate}
\item If the configuration $\vec{x}\in\XD^n$ is a local fixed point of 
	  $S$ for $U'\subseteq V$ and $\vec{x}$ is a local fixed point of $S$ 
	  for $U''\subseteq V$, then $\vec{x}$ is a local fixed point of $S$ for 
	  $U'\cup U''$.
\item A configuration $\vec{x}\in\XD^n$ is a local fixed point of $S$ for 
	  $U\subseteq V$ if and only if $\vec{x}$ is a local fixed point for 
	  all $U'\subseteq U$.
\end{enumerate}
\end{proposition}
\noindent
Notice that the second item of Proposition \ref{prop:fundamental} shows that 
the concept of fixed points is independent of update schedules.
\begin{corollary}
Let $S=(G,\{f_i~|~i\in V\})$ be a dynamical system over domain $\XD$. 
A configuration $\vec{x}\in\XD^n$ is a fixed point of $S$ if and only if 
for all update schedules $\alpha:\{1,\dots,T\}\to \pot(V)$, it holds that 
$\XF_{(S,\alpha)}(\vec{x})=\vec{x}.$ 
\end{corollary}

\section{The Analysis Framework}
\label{sec:analysis-framework}

In this section, we give a formal description of classification schemes 
for dynamical systems. 
Local transition functions are classified by Post classes (i.e., closures 
under composition) and networks are classified by forbidden minors (i.e., 
closures under taking minors).

\medskip
\noindent
{\em Post classes.} We adopt notation from \cite{boehler-creignou-reith-vollmer-2003}.
An $n$-ary boolean function $f$ is a mapping $f:\{0,1\}^n\to\{0,1\}$. Let $\BF$
denote the class of all boolean functions. 
There are two $1$-ary boolean functions: $\id(x)=_{\rm def} x$ and $\nt(x)=_{\rm def} 
1-x$ (which are denoted in formulas by $x$ for $\id(x)$ and $\overline{x}$ 
for $\nt(x)$). 
For $b\in\{0,1\}$, a boolean function $f$ is said to be {\em $b$-reproducing} 
if and only if $f(b,\dots,b)=b$. 
For binary $n$-tuples $\vec{a}=(a_1,\dots,a_n)$ and $\vec{b}=(b_1,\dots,b_n)$, 
we say that $(a_1,\dots,a_n)\le (b_1,\dots,b_n)$ if and only if for all $i\in \{1,\dots,n\}$,  
it holds that $a_i\le b_i$. An $n$-ary boolean function $f$ is said to be {\em monotone} 
if and only if for all $\vec{x},\vec{y}\in\{0,1\}^n$, $\vec{x}\le\vec{y}$ implies 
$f(\vec{x})\le f(\vec{y})$. 
An $n$-ary boolean function $f$ is said to be {\em self-dual} if and only if for all 
$(x_1,\dots,x_n)\in\{0,1\}^n$, it holds that $f(x_1,\dots,x_n)=\nt(f(\nt(x_1),\dots,
\allowbreak\nt(x_n)))$. 
A boolean function $f$ is linear if and only if there are constants 
$a_1,\dots,a_n\in\{0,1\}$ such that $f(x_1,\dots,x_n)=a_0\oplus a_1x_1\oplus \cdots \oplus 
a_nx_n$. 
Here, $\oplus$ is understood as addition modulo $2$ and $xy$ is understood as 
multiplication modulo $2$. 

We say that a function class $\F$ is {\em Post} if and only if $\F$ contains the function $\id$ and
$\F$ is closed under introduction of fictive variables, permutations of variables, identification of variables, and superposition (see, e.g., \cite{boehler-creignou-reith-vollmer-2003} for a formal definition).
It is a famous theorem by Post \cite{post-1941} that the family of all Post 
classes of boolean functions is a countable lattice with respect to set 
inclusion (see, e.g., \cite{post-1941,zverovich-2005} for a proof). The maximal 
meet-irreducible classes are the following (see, e.g., \cite{boehler-creignou-reith-vollmer-2003}):
\[\begin{array}{rll}
\BFR_0 
&=_{\rm def} \{~f\in\BF~|~\textrm{$f$ is $0$-reproducing}~\}
&\textrm{ with logical basis $\{ \AND, \XOR\}$}\\[.5ex]
\BFR_1 
&=_{\rm def} \{~f\in\BF~|~\textrm{$f$ is $1$-reproducing}~\}
&\textrm{ with logical basis $\{\OR, \XNOR\}$}\\[.5ex]
\BFL   
&=_{\rm def} \{~f\in\BF~|~\textrm{$f$ is linear}~\}
&\textrm{ with logical basis $\{\XOR, 0,1\}$}\\[.5ex]
\BFM   
&=_{\rm def} \{~f\in\BF~|~\textrm{$f$ is monotone}~\}
&\textrm{ with logical basis $\{\AND,\OR, 0,1\}$}\\[.5ex]
\BFD   
&=_{\rm def} \{~f\in\BF~|~\textrm{$f$ is self-dual}~\}
&\textrm{ with logical basis }\\
&
&\qquad\qquad \{(x\land\overline{y}) \vee 
								 (x\land\overline{z}) \vee 	
								 (\overline{y}\land\overline{z}) \}
\end{array}\]
From the structure of Post's lattice follows that each other class which does 
not contain all boolean functions is included in an intersection
of two of the classes listed.

\medskip

\noindent
{\em Graph minor classes.}
We adopt notation from \cite{diestel-2003}. 
Let $X$ and $Y$ be two undirected graphs. 
We say that $X$ is minor of $Y$ if and only if there is an isomorphic subgraph $Y'$ 
of $Y$ such that $X$ is obtained by contracting edges of $Y'$.
Let $\preceq$ be the relation on graphs defined by $X\preceq Y$ if and only 
if $X$ is a minor of $Y$. A class $\XG$ of graphs is said to be {\em closed 
under taking minors} if and only if for all graphs $G$ and $G'$, if $G\in\XG$ 
and $G'\preceq G$, then $G'\in \XG$. Let $\XX$ be any set of graphs. 
$\Forb_\preceq(\XX)$ denotes the class of all graphs without a minor in $\XX$ 
(and which is closed under isomorphisms). More specifically, 
$\Forb_\preceq(\XX)=_{\rm def} \{G~|~\textrm{$G\not\succeq X$ for all 
$X\in\XX$}~\}$. The set $\XX$ is called the set of {\em forbidden minors}.
Note that $\Forb_\preceq(\emptyset)$ is the class of all graphs.
As usual, we write $\Forb_\preceq(X_1,\dots,X_n)$ instead of 
$\Forb_\preceq(\{X_1,\dots,X_n\})$. A useful property of the forbidden-minor 
classes is the monotonicity with respect to $\preceq$, i.e., $X\preceq Y$ 
implies $\Forb_\preceq(X)\subseteq \Forb_\preceq(Y)$.

The celebrated Graph Minor Theorem, due to Robertson and 
Seymour \cite{robertson-seymour-2004}, shows that there are only countably
many graph classes closed under taking minors: A class $\XG$ of graphs is 
closed under taking minors if and only if there is a finite set $\XX$ such that 
$\XG=\Forb_\preceq(\XX)$. An important consequence of this theorem is that
for each graph class $\XG$ which is closed under taking minors there 
exists a polynomial-time algorithm to decide whether a given graph belongs to $\XG$
\cite{robertson-seymour-1995}.

The most prominent example of a characterization of a class closed under 
taking minors in terms of forbidden sets are the planar graphs. Let $K^n$ 
denote the complete graphs on $n$ vertices and let $K_{n,m}$ denote the 
complete bipartite graph having $n$ vertices in one component and $m$ 
vertices in the other component. The well-known Kuratowski-Wagner theorem
(see, e.g., \cite{diestel-2003}) states that a graph $G$ is 
planar if and only if $G$ belongs to $\Forb_\preceq(K_{3,3},K^5)$.
Planar graphs have an algorithmically important property: a graph $X$ is 
planar if and only if $\Forb_\preceq(X)$ has bounded treewidth \cite{robertson-seymour-1986}. 
As we use treewidth of a graph only in a black-box fashion, we refer to, 
e.g., \cite{diestel-2003} for a definiton of treewidth. 
A class $\XG$ of graphs is said to have {\em bounded treewidth} if and only if 
there is a $k\in\Nat$ such that all graphs in the class have treewidth at 
most $k$.

A similar but much less subtle behavior to planar graphs show graphs with a vertex 
cover of size one. Let $G=(V,E)$ be a graph. We say that a subset $U\subseteq V$ 
is a {\em vertex cover} of $G$ if and only if for all edges $\{u,v\}\in E$,
it holds that $\{u,v\}\cap U\not=\emptyset$. It is known that the class of graphs 
having a vertex cover of size at most $k$ is closed under taking minors 
\cite{cattell-dinneen-1994}. Moreover, $G$ has a vertex cover of size one if and 
only if $G$ belongs to $\Forb_\preceq(K^3,K^2\oplus K^2)$ \cite{cattell-dinneen-1994},
where for graphs $G$ and $G'$, $G\oplus G'$ denotes the graph obtained by the disjoint 
union of $G$ and $G'$. 
A class of graphs is said to have {\em bounded degree} if and only if there is
a $k\in\Nat$ such that all graphs in the class have a maximum vertex-degree of at
most $k$. 

\begin{proposition}
Let $X$ be a graph. Then, $X$ has a vertex cover of size one if and only if
$\Forb_\preceq(X)$ has bounded degree.
\end{proposition}

\begin{proof}
For $(\Rightarrow)$, suppose that $X$ has a vertex cover of size one. Then, $X$ 
consists of some star graph $K_{1,k}$ and some isolated vertices $u_1,\dots,u_r$.
Assume that $\Forb_\preceq(X)$ does not have bounded degree. Thus, there exists a
graph $G\in\Forb_\preceq(X)$ of maximum vertex-degree at least $k+r$. So, $G$ 
contains a subgraph $K_{1,k+r}$. Hence, $X\preceq G$, contradicting 
$G\in\Forb_\preceq(X)$. Therefore, $\Forb_\preceq(X)$ has bounded degree.

For $(\Leftarrow)$, suppose that $X$ does not have any vertex cover of size one.
It is easily seen that in this case, $X$ contains a triangle or two non-incident 
edges. First, suppose $X$ contains a triangle $K^3$. As $\Forb_\preceq(K^3)$
contains the class of all trees, which certainly does not have bounded degree,
we easily obtain from $K^3\preceq X$, that $\Forb_\preceq(X)$ does not have 
bounded degree as well. Second, suppose $X$ contains at least two non-incident 
edges, i.e., $K^2\oplus K^2\preceq X$. It follows that $\Forb_\preceq(X)$ contains
for all $k\in\Nat$, the star graph $K_{1,k}$. Thus, $\Forb_\preceq(X)$ does not 
have bounded degree. This completes the proof of the direction from left to right.
\end{proof}

\section{Islands of Tractability for Fixed-Point Existence}
\label{sec:islands}

In this section we are interested in the computational complexity of the following 
problem. Let $\F$ be a class of boolean functions and let $\XG$ be a class of graphs.

\bigskip

\noindent
\begin{tabular}{p{.14\linewidth}p{.8\linewidth}}
  \textsl{Problem:}  	& $\FPE(\F,\XG)$ \\
  \textsl{Input:}    	& A boolean dynamical system $S=(G,\{f_1,\dots,f_n\})$ 
						  such that $G\in\XG$ and for all $i\in\{1,\dots,n\}$, 
						  $f_i\in\F$\\
  \textsl{Question:}   	& Does $S$ have a fixed point?
\end{tabular}

\bigskip
\noindent
The complexity of the problem depends on how transition functions are represented.
We consider the cases of look-up table, formula, and circuit representations. 
The corresponding problems are denoted by $\FPE_{\lookup}$, $\FPE_\formula$, and
$\FPE_\circuit$. 
It is obvious that all problem versions belong to $\NP$. 
We say that a problem is intractable if it is $\NP$-hard, and it is tractable if it 
is solvable in polynomial time.

\subsection{The Case of Local Transition Functions Given by Look-up Tables}

The main result of this subsection is the following dichotomy result. 

\begin{theorem}\label{thm:fpe-dichotomy-t}
\sloppy Let $\F$ be a Post class of boolean functions and let 
$\XG$ be a class of graphs closed under taking minors. If $\F\supseteq \BFD$ 
and $\XG\supseteq \Forb_\preceq(K_{3,3},K^5)$ then $\FPE_{\lookup}(\F,\XG)$ 
is intractable. Otherwise, $\FPE_{\lookup}(\F,\XG)$ is tractable.
\end{theorem}

We postpone the proof to the end of this paragraph when we have proved a number
of complexity results for several classes of systems.

\medskip

\noindent
{\em Dynamical systems with a tractable fixed-point analysis.}
We first summarize the tractable classes of local transition functions of systems 
on arbitrary underlying networks. The results presented next are 
well-known or follow easily from the definitions.
\begin{proposition}\label{prop:tractable-transitions}
\begin{enumerate}
\item $\FPE_{\lookup}(\BFR_1,\Forb(\emptyset))$ is solvable in polynomial time.
\item $\FPE_{\lookup}(\BFR_0,\Forb(\emptyset))$ is solvable in polynomial time.
\item {\rm \cite{barrett-hunt-marathe-ravi-rosenkrantz-stearns-tosic-2001}}
	  $\FPE_{\lookup}(\BFM,\Forb(\emptyset))$ is solvable in polynomial time.
\item {\rm \cite{barrett-hunt-marathe-ravi-rosenkrantz-stearns-tosic-2001}}
	  $\FPE_{\lookup}(\BFL,\Forb(\emptyset))$ is solvable in polynomial time.
\end{enumerate}
\end{proposition}
\noindent
Notice that $\BFD$ is the only remaining class which does not contain all boolean
functions.

Next we restrict network classes. To identify tractable cases, we express the 
fixed-point existence problem as a constraint satisfaction problem.
A constraint satisfaction problem (CSP) consists of triples $(X,\XD,\XC)$, where 
$X=\{x_1,\dots,x_n\}$ is the set of variables, $\XD$ is the domain of the variables,
$\XC$ is a set of constraints $Rx_{i_1},\dots,x_{i_k}$ having associated corresponding
relations $R_{{i_1},\dots,{i_k}}$, such that there exists an assignment $I:X\to\XD$ 
satisfying $(I(x_{i_1}),\dots,I(x_{i_k}))\in R_{{i_1},\dots,{i_k}}$ for all constraints
$Rx_{i_1},\dots,x_{i_k}\in \XC$. We suppose that $\XC$ is listed by pairs 
$\langle Rx_{i_1},\dots,x_{i_k},\allowbreak R_{{i_1},\dots,{i_k}} \rangle$.

\begin{theorem} \label{thm:tractable-networks}
Let $X$ be a planar graph. Then, $\FPE_{\lookup}(\BF,\Forb_\preceq(X))$ is solvable in polynomial time.
\end{theorem}

\begin{proof}
We describe a general reduction from $\FPE_{\lookup}(\F,\XG)$ to constraint satisfaction 
problems. Suppose we are given a dynamical system $S=(G,\{f_1,\dots,f_n\})$. 
Let $G=(V,E)$ be the underlying network. Without loss of generality, we may 
assume that $G$ does not have isolated vertices, i.e., $d_i\ge 1$ for all 
$i\in V$. Define $\CSP(S)=(X,\XD,\XC)$ to be the constraint satisfaction 
problem specified as follows:
\begin{eqnarray*}
X 	&=_{\rm def}& \{x_1,\dots,x_n\} \\[1ex]
\XD &=_{\rm def}& \textstyle \bigcup_{i\in V} \XD_i \textrm{ where for all $i\in V$,}\\
	& 			& \XD_i=_{\rm def} \{(I,i)~|~\textrm{$I:N_G^0(i)\to\{0,1\}$
								such that $f_i(I(i_0),\dots,I(i_k))=I(i)$}\\
	& 			& \qquad\qquad\qquad\quad\textrm{where $\{i_0,\dots,i_k\}=N_G^0(i)$}\}\\[1ex]
\XC &=_{\rm def}& \{ Ex_ix_j~|~\{i,j\}\in E(G)\textrm{ and } i\le j\} \textrm{ where for all $i\le j$,}\\
    & 			& E_{ij}=_{\rm def}\{~ \{(I_i,i), (I_j,j)\} ~|~
					\textrm{ $I_i(k)=I_j(k)$ for all $k\in N_G^0(i)\cap N_G^0(j)$} ~\}
\end{eqnarray*}
Let $n=\|V\|$ and $m=\|E\|$. By construction of $\CSP(S)$ we obtain that $\|X\|=n$ 
and that the number of constrains in $\XC$ is just $m$. 
The size of the domain $\D$ is at most proportional to $\sum_{i\in V} (1+d_i)\cdot 2^{1+d_i}$, 
and the size of the set of constraint relations can bounded by
\[\textstyle c\cdot \sum_{\{i,j\}\in E} (2+d_i+d_j)\cdot 2^{1+d_i}\cdot 2^{1+d_j}
\le c\cdot \left(\sum_{i\in V} (1+d_i)\cdot 2^{1+d_i}\right)^2\]
for some constant $c>0$. The latter holds because $d_i\ge 1$ for all $i\in V$. All in all, this easily implies that 
$|\CSP(S)|=O(|S|^2)$. Hence, $\CSP(S)$ is computable in polynomial time in the size of $S$.
We have to show that
\[\textrm{$S$ has a fixed point $~~\Longleftrightarrow~~$ 
$\CSP(S)$ has a satisfying assignment.}\]
We prove both directions of the equivalence separately.

For $(\Rightarrow)$, suppose the configuration $I:V\to\{0,1\}$ is a fixed point
of $S$.  Define an assigment $I':\{x_1,\dots,x_n\}\to \XD$ by
$I'(x_i)=_{\rm def} I[N_G^0(i)]$.
We show that $I'$ satisfies all constraints. For all $i\in V$, let $I_i$ denote 
$I[N_G^0(i)]$. Let $Ex_ix_j$ be any constraint of $\CSP(S)$ and let $E_{ij}$ be 
the relation associated with $Ex_ix_j$. By definition, $\{i,j\}\in E$. 
Since $I$ is a fixed point, it holds that $(I_i,i)$ and $(I_j,j)$ belong to 
$\XD$. For all $k\in N_G^0(i)\cap N_G^0(j)$ we obtain 
$I_i(k)=I'(x_i)(k)=I'(x_j)(k)=I_j(k)$. Thus, $\{ (I_i,i),(I_j,j)\}$ lies in 
$E_{ij}$. This proves the direction from left to right.

For $(\Leftarrow)$, suppose $I$ is a satisfying assignment for $\CSP(S)$. Let 
$(I_i,i)\in \XD$ be the pair which $I$ assigns to the variable $x_i\in V$. 
Define a configuration $I':V\to\{0,1\}$ by
$I'(i) =_{\rm def} I_i(i).$
We are done if we are able to show that the following is true for all $i\in V$:
\begin{equation}
I'[N_G^0(i)]\equiv I_i \label{eqn:compatible-projection}
\end{equation}
Since for all $i\in V$, $f_i(I_i(i_0),\dots,I_i(i_k))=I_i(i)$ where 
$\{i_0,\dots,i_k\}=N_G^0(i)$, Eq.~(\ref{eqn:compatible-projection}) implies 
that $I'$ is a fixed point of $S$. To verify the equation, we consider $i$ and 
all neighbors of $i$ individually. Let $j\in N_G^0(i)$. If $j=i$, there is 
nothing to show. So let $j\not=i$. Since $\{i,j\}\in E(G)$ and since $I$ is 
a satisfying assignment, $I_i(k)=I_j(k)$ holds for all 
$k\in N_G^0(i)\cap N_G^0(j)$. Thus, we obtain  $I'(j) = I_j(j) = I_i(j)$.
This shows the correctness of Eq.~(\ref{eqn:compatible-projection}) 
for all $i\in V$. Hence, the direction from right to left is proved.

We thus have established a polynomial-time many-one reduction between 
the fixed-point existence problem and constraint satisfaction problems. 
Moreover, it is easy to see that the graph $G$ of a given dynamical system 
$S$ is isomorphic to $\CSP(S)$'s constraint graph which consists of the vertex 
set $\{x_1,\dots,x_n\}$ and the edge set $\{\{x_i,x_j\} ~|~ Ex_ix_j\in \XC\}$. 
Thus, if $X$ is planar, then $\Forb_\preceq(X)$ has bounded treewidth, and so
the constraint graph of $\CSP(S)$ has bounded treewidth for each dynamical 
system $S=(G,\{f_1,\dots,f_n\})$ such that $G\in\Forb_\preceq(X)$. The theorem 
follows from the well-known polynomial-time algorithms for constraint 
satisfaction problems having constraint graphs of bounded treewidth 
\cite{freuder-1990}.
\end{proof}

\noindent
{\em Dynamical systems with an intractable fixed-point analysis.}
We turn to the intractable cases of the fixed-point existence problem. 
As a first step, $\FPE_{\lookup}\allowbreak(\F,\XG)$ is shown to be $\NP$-complete for arbitrary 
boolean, local transition functions and planar networks. Recall that similar
results have only been shown for formula respresentation.

We use $\PSAT$ to prove $\NP$-hardness. Let $H=C_1\land\cdots\land C_m$ be
a propositional formula in conjuctive normal form where each clause $C_j$ 
consists of three literals (for short, 3CNF) where positive literals $x_i$ and 
negative literals $\neg x_i$ are taken from the set of variables $\{x_1,\dots,x_n\}$. 
The graph representation $\Gamma(H)$ of $H$ is a bipartite graph consisting of the 
vertex set $\{x_1,\dots,x_n, C_1,\dots,C_m\}$ and all edges $\{x_i,C_j\}$ such 
that variable $x_i$ appears as a literal in the clause $C_j$. A planar 3CNF
is a 3CNF such that its graph representation is planar. $\PSAT$ is the problem
to decide whether a given planar 3CNF is satisfiable. It is well known that
$\PSAT$ is an $\NP$-complete problem \cite{lichtenstein-1982}.

\begin{theorem}\label{thm:fpe-bf-planar}
$\FPE_{\lookup}(\BF,\Forb(K_{3,3}, K^5))$ is $\NP$-complete, even restricted to underlying 
networks having maximum vertex degree three.
\end{theorem}

\noindent
Note that for graphs having maximum degree at most two the problem is tractable.
\begin{proof}
Let $H=C_1\land C_2\land\dots\land C_m$ be a 3CNF having variables $x_1,\dots,
x_n$ and a planar graph representation $\Gamma(H)$ where $V(\Gamma(H))=U\cup W$,
$U=\{x_1,\dots,x_n\}$, and $W=\{C_1,\dots,C_m\}$. Suppose that $V(\Gamma(H))$ is
totally ordered such that the clause vertices of $W$ come before the variable vertices
of $U$.

We construct the following system. For the underlying network $G=(V,E)$,
compute an embedding of $\Gamma(H)$ in the plane (in linear time). Now replace 
the vertices of $U$ in the following way. Let $x_i\in U$ and suppose that 
$x_i$'s neighbors $C_{j_1},\dots,C_{j_r}$ in $\Gamma(H)$ are clockwise ordered 
with respect to the planar embedding, such that $j_1$ is minimal in 
$\{j~|~\textrm{$x_i$ appears as a literal in $C_j$}~\}$.  Replace the vertex 
$x_i\in U$ by a cycle 
$\{x_{i,j_1},\dots,x_{i,j_r}\}$ which is connected to the neighbors of $x_i$ 
in $\Gamma(H)$ by having an edge $\{x_{i,j},C_j\}$ whenever $x_i$ appears as a 
literal in $C_j$.  Let $G=(V,E)$ be the graph obtained after all replacements 
are made in this way. Clearly, $G$ is planar, i.e., $G\in \Forb_\preceq(K_{3,3},
K^5)$, and can be computed in time polynomial in the size of $H$. Note that the 
maximum degree of a vertex in $G$ is three.
To complete the construction, we specify the local transition functions.
Let $C_i\in W$ be clause vertex. Let $x_{i_1,j_1},x_{i_2,j_2}$, and $x_{i_3,j_3}$
be the neighbors of $C_i$ in $G$. Define the local transition function $f_{C_i}$ by
\begin{eqnarray*}
\lefteqn{f_{C_i}(I(C_i), I(x_{i_1,j_1}),I(x_{i_2,j_2}),I(x_{i_3,j_3}))}\\[0.5ex]
& & =_{\rm def}\left\{\begin{array}{ll}
1	 	&\textrm{ if $\bigl(I(x_{i_1,j_1}),I(x_{i_2,j_2}),I(x_{i_3,j_3})\bigr)$ is
				a satisfying assignment of $C_i$}\\
\nt(I(C_i)) &\textrm{ otherwise}
\end{array}\right.
\end{eqnarray*}
Let $x_{i,j}\in V$ be a variable vertex. Suppose that $\{C_j, x_{i,k_0},\dots,
x_{i,k_r}\}=N^0_G(x_{i,j})$ where $r$ is the degree of 
$x_{i,j}$ in $G$. Define the local transition function $f_{x_{i,j}}$ by
\[f_{x_{i,j}}(I(C_j),I(x_{i,k_0}),\dots, I(x_{i,k_r}))
=_{\rm def}\left\{\begin{array}{ll}
I(x_{i,j}) &\textrm{ if $I(x_{i,k_0})=\dots=I(x_{i,k_r})$}\\
\nt(I(x_{i,j})) &\textrm{ otherwise}
\end{array}\right.\]
Let $S_H$ denote the dynamical system $(G, \{f_v|v\in V\}$ constructed 
from any planar 3CNF $H$ in the way just specified. Note that, since the 
maximum degree of any vertex in $G$ is three, we can compute 
$S_H$ in time polynomial in the size of $H$. 

By construction of $S_H$, a configuration $I$ is a fixed point of $S_H$
if and only if $I$ satisfies $I(x_{i,j_1})=\dots=I(x_{i,j_r})$ for all 
$i\in\{1,\dots,n\}$ and, furthermore, $I(C_j)=1$ for all $j\in\{1,\dots,m\}$. 
Hence, it is easily seen that $H$ has a satisfying assignment if and only 
if $S_H$ has a fixed point.
\end{proof}

We extend the $\NP$-completeness of the fixed-point existence problem to 
systems with self-dual transition functions and planar graphs. First, 
observe that the transition functions for clause variables are not self-dual.
This implies that the construction used in the last theorem is not fully
appropriate. However, arbitrary boolean functions 
can easily be embedded into a self-dual function of larger arity.

\begin{proposition}\label{prop:selfdualizer}
Let $n\in\Nat_+$. For each $k$-ary boolean function $f:\{0,1\}^k\to\{0,1\}$,
the function $\sd_n(f):\{0,1\}^{k+n+1}\to\{0,1\}$ defined for all
$x_1,\dots,x_k,y_1,\dots, y_n, z\in\{0,1\}$ by
\[
\sd_n(f)(x_1\dots,x_k,y_1,\dots y_n,z)
=_{\rm def} \left\{\begin{array}{ll}
f(x_1,\dots,x_k) 	 & \textrm{if $y_1=\dots=y_n=0$}\\
\nt(f(\nt(x_1),\dots,\nt(x_k))) 
					 & \textrm{if $y_1=\dots=y_n=1$}\\
\nt(z)				 & \textrm{otherwise}
\end{array}\right.
\]
is self-dual.
\end{proposition}

\begin{proof}
Follows from the definitions by case analysis.
\end{proof}

The usage of Proposition \ref{prop:selfdualizer} introduces ambiguity to
the set of fixed points. 

\begin{proposition}\label{prop:fp-mirroring}
Let $S=(G,\{f_i~|~i\in V\})$ be a dynamical system over 
$\{0,1\}$ so that all local transition functions $f_i$ are self-dual. Let 
$U\subseteq V$. Then, a configuration $I:V\to\{0,1\}$ is a local fixed point
of $S$ for $U$ if and only if the configuration $\overline{I}:V\to\{0,1\}$ defined by 
$\overline{I}(i)=\nt(I(i))$, is a local fixed point of $S$ for $U$.
\end{proposition}

\begin{proof}
Immediate from the definitions.
\end{proof}

\begin{theorem} \label{thm:fpe-d-planar}
$\FPE_{\lookup}(\BFD,\Forb(K_{3,3},K^5))$ is $\NP$-complete.
\end{theorem}

\begin{proof}
We reduce from $\FPE_{\lookup}(\BF,\Forb(K_{3,3},K^5))$. Let $S=(G,\{f_i~|~i\in V\})$ be a 
dynamical system such that the underlying network $G$ is planar. We construct another
system $S'=(G',\{f'_i~|~i\in V'\})$ as follows. The network $G'=(V',E')$ is defined by
$V'=_{\rm def} V\cup E$ and $E'=_{\rm def} E\cup \{~\{i,e\}~|~i\in V,e\in E, i\in e~\}$.
It is easily seen that $G'$ is planar as well. Suppose $V'$ is ordered such that $V$ is 
in the same ordering as for $S$, $E$ is arbitrarily ordered, and $E$ comes completely 
after $V$. The local transition functions of the vertices of $V'$ are specified as follows. 
Suppose $i\in V$. Let $\{i_0,\dots,i_k\}=N^0_{G'}(i)\cap V$ and let
$\{e_1,\dots,e_k\}=N^0_{G'}(i)\cap E$. We define the local transition 
function $f'_i$ by
\[f'_i(x_{i_0},\dots,x_{i_k},x_{e_1},\dots,x_{e_k})=_{\rm def}
\sd_k(f_i)(x_{i_0},\dots,x_{i_k},x_{e_1},\dots,x_{e_k},x_i).\]
By Proposition \ref{prop:selfdualizer} and since $\BFD$ is closed under identification 
of variables, $f'_i$ is self-dual. Moreover, as the degree of $i$ in $V'$ is $2k$ where 
$k$ is the degree of $i$ in $G$, the function table can be computed in polynomial time. 
Now consider a vertex $e=\{i,j\}\in E$. Define the local transition function $f'_e$ by
\[f'_{\{i,j\}}(x_i,x_j,x_{\{i,j\}})=_{\rm def} x_{\{i,j\}}.\]
Clearly, $f'_e$ is self-dual. As the degree of $e\in E$ is two, the function table is trivially
computable in polynomial time. 
It remains to show that 
\[\textrm{$S$ has a fixed point}~~\Longleftrightarrow~~\textrm{$S'$ has a fixed point.}\]
We prove both direction individually.

For $(\Rightarrow)$, suppose the configuration $I:V\to\{0,1\}$ is a fixed point of $S$, 
i.e., for all $i\in V$, it holds that $f_i(I(i_0),\dots,I(i_k))=I(i)$ where 
$\{i_0,\dots,i_k\}=N^0_G(i)$.  Define a configuration $I':V'\to\{0,1\}$ for 
$i\in V$ by $I'(i)=_{\rm def} I(i)$ and for $e\in E$ by $I'(e)=_{\rm def} 0$. Consider 
$i\in V$. Assume $\{i_0,\dots,i_k\}=N^0_{G'}(i)\cap V$ and $\{e_1,\dots,e_k\}=
N^0_{G'}(i)\cap E$. We obtain
\begin{eqnarray*}
\lefteqn{f'_i(I'(i_0),\dots,I'(i_k),I'(e_1),\dots,I'(e_k))}\\
&=&\!\!\! \sd_k(f_i)(I(i_0),\dots,I(i_k),0,\dots,0,I(i))
= f_i(I(i_1),\dots,I(i_k))
= I(i)
= I'(i).
\end{eqnarray*}
Thus, $I'$ is a local fixed point for $i\in V$. Suppose $e=\{i,j\}\in E$. 
By definition, $f'_{\{i,j\}}(I'(i),I'(j),I'(\{i,j\}))=I'(\{i,j\})$. Thus, $I'$
is a local fixed point for $e\in E$. Proposition \ref{prop:fundamental} implies 
that $I'$ is a fixed point of $S'$.

For $(\Leftarrow)$, suppose the configuration $I':V'\to\{0,1\}$ is a fixed point
of $S'$. Observe that  for all $e,e'\in E\subseteq V'$, there is a walk in $G'$ 
from $e$ to $e'$ alternating between vertices in $V$ and $E$, i.e., there are 
vertices $p_0,\dots,p_{2\ell}\in V'$ such that $\{p_i,p_{i+1}\}\in E'$ for all 
$0\le i<2\ell$, $p_0=e$, $p_{2\ell}=e'$, and for all $0\le j<\ell$, it holds 
that $p_{2j}\in E$ and $p_{2j+1}\in V$. Consider a vertex $p_{2j+1}\in V$. 
Let $\{i_0,\dots,i_k\}=N^0_{G'}(p_{2j+1})\cap V$ and $\{e_1,
\dots,e_k\}=N^0_{G'}(p_{2j+1})\cap E$. Since $I'$ is a fixed 
point of $S$, we have
\begin{eqnarray*}
I'(p_{2j+1}) 
&=& f'_{p_{2j+1}}(I'(i_0),\dots,I'(i_k), I'(e_1),\dots, I'(e_k))\\
&=& \sd_k(f_{p_{2j+1}}(I'(i_0),\dots, I'(i_k), I'(e_1),\dots, I'(e_k), I'(i))
\end{eqnarray*}
By definition of $\sd_k(f_{p_{2j+1}})$, it follows that $I'(e_1)=\dots=I'(e_k)$. 
In particular, $I'(p_{2j})=I'(p_{2j+2})$. This implies that for all $e,e'\in E$, 
it holds that $I'(e)=I'(e')$. By Proposition \ref{prop:fp-mirroring}, we may 
assume that $I'(e)=0$ for all $e\in E$. Define a configuration $I:V\to\{0,1\}$
of $S$ by $I(i)=I(i)$ for all $i\in V$.  It is easily seen that $I$ is a 
fixed point of $S$.
\end{proof}

\medskip
\noindent
{\em Composing the big picture.} We come back to the proof of the main 
result of the subsection. For convenience, we state it once more.

\bigskip
\noindent
{\bf Theorem {\bf \ref{thm:fpe-dichotomy-t}}.}
{\em Let $\F$ be a Post class of boolean functions and let 
$\XG$ be a class of graphs closed under taking minors. If $\F\supseteq \BFD$ 
and $\XG\supseteq \Forb_\preceq(K_{3,3},K^5)$ then $\FPE_{\lookup}(\F,\XG)$ 
is intractable. Otherwise, $\FPE_{\lookup}(\F,\XG)$ is tractable.}

\begin{proof}
If $\F\supseteq \BFD$ and $\XG\supseteq\Forb(K_{3,3},K^5)$, then $\FPE(\F,\XG)$
is $\NP$-complete by Theorem \ref{thm:fpe-d-planar}. Suppose $\F\not\supseteq 
\BFD$ or $\XG\not\supseteq\Forb(K_{3,3},K^5)$. The maximal classes $\F$ that do
not contain $\BFD$ are $R_1$, $R_0$, $M$, and $L$. For all these classes, by
Proposition \ref{prop:tractable-transitions}, the fixed-point existence problem
is solvable in polynomial time. It remains to consider the case that $\XG\not
\supseteq\Forb(K_{3,3},K^5)$. Suppose $\XG=\Forb(X_1,\dots,X_n)$. 
Since $\XG\not\supseteq \Forb_\preceq(K_{3,3},K^5)$, there is an $i$ such that 
$X_i$ is planar. Since $\XG\subseteq \Forb_\preceq(X_i)$, Theorem \ref{thm:tractable-networks}
shows that $\FPE_\lookup(\BF,\XG)$ is solvable in polynomial time.
\end{proof}

\subsection{Succinctly Represented Local Transition Functions}

In this section we prove a dichotomy theorem for the fixed-point existence 
problem when transitions are given by formulas or circuits. As usual, the size 
of formula is the number of symbols from the basis used to encode the formula, 
the size of a circuit is the number of gates it consists of (including the 
input gates). Both succinct representations of functions lead to the same result.

\begin{theorem}\label{thm:fpe-dichotomy-f}
Let $\F$ be a Post class of boolean functions and let $\XG$ be a class of 
graphs closed under taking minors. 
If $\F\supseteq\BFD$ and $\XG\supseteq\Forb_\preceq\allowbreak(K^3, K^2\oplus K^2)$, 
 then $\FPE_\formula\allowbreak(\F,\XG)$ is intractable.
Otherwise, $\FPE_\formula(\F,\XG)$ is tractable.
Moreover, the same statement holds for $\FPE_\circuit$.
\end{theorem}
\noindent
Again we postpone the proof until we have proved a number of special results. 

On the side of tractable cases, first note that Proposition 
\ref{prop:tractable-transitions} still holds for formulas and circuits. Actually, 
the results in \cite{barrett-hunt-marathe-ravi-rosenkrantz-stearns-tosic-2001} 
were stated for formulas. Furthermore, notice that circuits over the basis 
$\{\XOR,0,1\}$ can easily be transformed in polynomial time into equivalent 
formulas over the same basis. The following result provides the tractability
limit for restricted network classes.

\begin{theorem}\label{thm:fpe-bf-bdegree}
Let $X$ be a graph having a vertex cover of size one. Then, 
$\FPE_\formula\allowbreak(\BF,\Forb_\preceq(X))$ and $\FPE_\circuit(\BF,\Forb_\preceq(X))$
are solvable in polynomial time.
\end{theorem}

\begin{proof}
Suppose $X$ has a vertex cover of size one. Then, there is an $r\in\Nat$ 
such that all graphs in $\Forb_\preceq(X)$ have maximum vertex degree $r$. Thus, 
if we compute lookup tables from formulas or circuits, then each lookup table has 
at most $2^{r+1}$ entries. Hence, in polynomial time, we can transform each 
dynamical system $S$ with a network in $\Forb_\preceq(X)$ and local transitions 
functions given by formulas or circuits into a dynamical system $S'$ with the 
same networks and local transition functions given by lookup tables 
such that $S$ and $S'$ have the same fixed-point configurations. Moreover, since 
$X$ is planar (note that $K^3\preceq K_{3,3}$ and $K^2\oplus K^2\preceq K_{3,3}$ as well as
$K^3\preceq K_{3,3}$ and $K^2\oplus K^2\preceq K^5$), Theorem \ref{thm:tractable-networks} implies polynomial-time
solvability of $\FPE_\formula(\BF,\Forb_\preceq(X))$ and 
$\FPE_\circuit(\BF,\Forb_\preceq(X))$.
\end{proof}

\begin{theorem}\label{thm:fpe-d-star}
$\FPE_\formula(\BFD,\Forb_\preceq(K^3, K^2\oplus K^2))$ is $\NP$-complete.
\end{theorem}

\begin{proof}
The reduction is from $\SAT$. We start with a description of a reduction to 
dynamical systems where each transition function is computed by a 3CNF. 
Suppose we are given a 3CNF 
$H=C_1\land\dots\land C_m$ having variables $x_1,\dots,x_n$. Note that the 
formula $H'=_{\rm def} H\vee \overline{x_0}$,  where $x_0$ is a new variable,
satisfies for any assignment $I:\{x_0,x_1,\dots,x_n\}\to
\{0,1\}$ that $f_{H'}(I(x_0),I(x_1),\dots,I(x_n))=I(x_0)$ if and only if 
$I(H)=1$ and $I(x_0)=1$. Moreover, $H'$ is equivalent to a 4CNF which can 
be transformed in a 3CNF $\hat{H}$ with variables $x_0,x_1,\dots,x_n,x_{n+1},
\dots,x_{n+m}$ such that $\hat{H}$ is satisfiable with $1$ assigned to 
$x_0$ if and only if $H$ is satisfiable with $1$ assigned to $x_0$. Define 
$S_H$ to be the dynamical system $(G,\{f_0,\dots,f_{n+m}\})$ consisting of 
the network $G=(V,E)$ where $V=_{\rm def} \{0,1,\dots,n+m\}$
and $E=_{\rm def} \{~\{0,i\}~|~i\in\{1,\dots,n+m\}~\}$.
Thus, $G$ is a star $K_{1,n+m}$, i.e., $G\in\Forb_\preceq(K^3, K^2\oplus K^2)$. 
The local transition functions are given as follows. For a vertex 
$i\in\{1,\dots,n+m\}$, the local transition function $f_i$ is defined
to be computed by the formula $H(x_0,x_i)=_{\rm def} x_i$.  For the central 
vertex, the local transition 
function $f_0$ is given by the formula $\hat{H}(x_0,x_1,\dots,x_{n+m})$ 
where the variable $x_i$ stands for a vertex $i\in V$. Clearly, $S_H$ can be 
computed in time polynomial in the size of $H$ and we have that $H$ is 
satisfiable if and only if $S_H$ has a fixed point.

We now transform the dynamical system $S_H$ into another system $S'_H$ with 
self-dual local transition functions given by formulas over the corresponding 
basis. Note that $\BFD$ has a single basis function of arity three. Let $D$ 
denote the corresponding ternary function symbol, i.e., the semantics of $D$ 
is defined by $D(x,y,z)\equiv_{\rm def} (x\land\overline{y})\vee 
(x\land\overline{z}) \vee (\overline{y}\land \overline{z})$. 
Note that $D(x,x,y)\equiv \overline{y}$. We embed a 3CNF into a self-dual 
function, similarly to Proposition \ref{prop:selfdualizer}. That is, for 
an arbitrary 3CNF $H=C_1\land\dots\land C_m$ 
having variables $x_1,\dots,x_n$, define the formula $\dual(H)(x_1,\dots,x_n)=
\overline{H(\overline{x_1}, \dots, \overline{x_n})}$. Define $\sd(H)(x_1,\dots,x_n,z) 
=_{\rm def} (H\land z) \vee (\dual(H)\land\overline{z})$. By induction over the 
number of clauses, we show that $\sd(H)$ is equivalent to a formula built using 
$D$ which is of polynomial size:
\begin{enumerate}
\item For the base of induction, suppose $m=1$. Since we know how to express 
	  negation using $D$, without loss of generality, we assume that 
	  $H=(x_1\vee x_2\vee x_3)$. So, $\sd(H)\equiv \bigl((x_1\vee x_2\vee x_3)
	  \land z\bigr) \vee \bigl((x_1\land x_2\land x_3)\land\overline{z}\bigr)$. 
	  By truth-table inspection we obtain that 
	  $\sd(H)\equiv\overline{D(\overline{z}, D(z,\overline{x_1},\overline{x_2}),
	  D(z,\overline{x_1},\overline{x_3}))}$.
\item For the induction step, suppose $m>1$. Let $H'=C_1\land\dots\land C_m$
	  be a 3CNF over the variables $x_1,\dots,x_n$. Define $H'=_{\rm def} 
	  C_1\land\dots\land C_{\lfloor m/2\rfloor}$ and $H''=_{\rm def} 
	  C_{\lfloor m/2\rfloor+1}\land\dots\land C_m$. Some equivalent 
	  transformations show that $\sd(H)\equiv D(\overline{z}, \overline{\sd(H')}, 
	  \overline{\sd(H'')})$. By induction hypothesis, $\sd(H')$ and $\sd(H'')$ 
	  can be expressed using $D$. Replacing $\sd(H')$ and $\sd(H'')$ (and the 
	  negation) gives the appropriate formula for $\sd(H)$. Note that the 
	  recursion depth for formula replacement is logarithmic. It follows
	  that the size of the formula for $\sd(H)$ is $O(|H|^2)$.
\end{enumerate}
Finally, we define the dynamical system $S'_H$ to be specified 
as follows. The network $G'=(V',E')$ consists of the vertex set
$V'=_{\rm def}\{\underline{0},0,\dots,n+m\}$ and the edge set 
$E'=_{\rm def} \{ \{\underline{0},i\}~|~i\in\{0,\dots,n+m\}~\}$.
Thus, $G'$ is a star $K_{1,n+m+1}$, i.e., $G'\in\Forb_\preceq(K^3, K^2\oplus K^2)$.
The local transition function $f_i$ for a vertex $i\in\{0,\dots,n+m\}$ 
is given by $H(x_{\underline{0}},x_i)=D(x_{\underline{0}},x_{\underline{0}},
D(x_{\underline{0}},x_{\underline{0}},x_i))$. Notice that 
$H(x_{\underline{0}},x_i)\equiv x_i$. The local transition function
$f_{\underline{0}}$ for the vertex $\underline{0}$ is given as follows. 
Recall that $\hat{H}$ is the 3CNF associated with the local transition function
of vertex $0$ in the system $S_H$. Then $f_{\underline{0}}$ is represented
by the $D$-formula equivalent to $\sd(\hat{H})(x_0,\dots,x_{n+m},
x_{\underline{0}})$. Clearly, $S'_H$ can be computed in time polynomial in 
the size of $H$. Moreover, it is easy to verify that $H$ is satisfiable if and 
only if $S'_H$ has a fixed point.
\end{proof}

We combine Theorem \ref{thm:fpe-bf-bdegree} and Theorem \ref{thm:fpe-d-star}
to prove Theorem \ref{thm:fpe-dichotomy-f}.

\begin{proof}{\em (Theorem \ref{thm:fpe-dichotomy-f})}
If $\F\supseteq \BFD$ and $\XG\supseteq\Forb(K^3, K^2\oplus K^2)$, then $\FPE(\F,\XG)$
is $\NP$-complete by Theorem \ref{thm:fpe-d-star}. Suppose $\F\not\supseteq 
\BFD$ or $\XG\not\supseteq\Forb(K^3, K^2\oplus K^2)$. The maximal classes $\F$ that do
not contain $\BFD$ are $R_1$, $R_0$, $M$, and $L$. For all these classes, by
Proposition \ref{prop:tractable-transitions}, the fixed-point existence problem
is solvable in polynomial time. It remains to consider the case that $\XG\not
\supseteq\Forb(K^3, K^2\oplus K^2)$. Suppose $\XG=\Forb(X_1,\dots,X_n)$. 
Since $\XG\not\supseteq \Forb_\preceq(K^3, K^2\oplus K^2)$, there is an $i$ such that 
$X_i$ has a vertex cover of size one. Since $\XG\subseteq \Forb_\preceq(X_i)$, 
Theorem \ref{thm:fpe-bf-bdegree} shows that $\FPE_\lookup(\BF,\XG)$ 
is solvable in polynomial time.
\end{proof}

\section{Conclusion}

We characterized the islands of tractability for the fixed-point 
existence problem for boolean dynamical systems with respect to transition classes $\F$
closed under composition and network classes $\XG$ closed under taking minors: If $\F$
contains the self-dual functions and $\XG$ contains the planar graphs, then 
$\FPE_\lookup(\F,\XG)$ is intractable, otherwise it is tractable. Replacing ``planar 
graphs'' by ``graphs having a vertex cover of size one'' yields the same dichotomy 
theorem for the succinct representations of local transition functions by formulas 
or circuits. The linear and monotone functions have been shown to be tractable cases in \cite{barrett-hunt-marathe-ravi-rosenkrantz-stearns-tosic-2001}. There, the authors suggested to 
find more tractable classes of local transition functions. Over the boolean domain our 
results show that, aside from two obvious exceptions (the $0$- and $1$-reproducing 
functions), there are no more such function classes.

Although the proposed analysis framework allows elegant dichotomy theorems 
for fixed-point existence, it is certainly necessary to examine its usefulness for 
other computational problems for discrete dynamical systems and to refine it appropriately.  
Another important open problem is the extension of the dichotomy theorems to larger
domains. This seems a challenging issue as even in the case of a ternary domain 
the number of Post classes is continuum (see, e.g., \cite{kaluzhnin-poeschel-1979}).

\bigskip
\noindent
{\bf Acknowledgements.}
I am grateful to Ernst W. Mayr (TU M\"unchen) for careful proofreading and pointing out an
error in earlier versions of this paper.

\end{document}